\documentclass[%
twocolumn,superscriptaddress,
 amsmath,amssymb,
 aps,
prc,
]{revtex4-2}

\usepackage{graphicx}
\usepackage{dcolumn}
\usepackage{bm}
\usepackage{hyperref}
\usepackage[mathlines]{lineno}
\usepackage{subeqnarray}
\usepackage{cases}
\usepackage{times}
\usepackage{datetime}
\usepackage{longtable}

\newcommand{\rr} {\boldsymbol{r}}

\begin{document}
\title{
High Quality Microscopic Nuclear Masses of Superheavy Nuclei 
}
\author{Dawei Guan}
\affiliation{
State Key Laboratory of Nuclear Physics and Technology, School of Physics,
Peking University, Beijing 100871, China
}
\author{Junchen Pei}\email{peij@pku.edu.cn}
\affiliation{
State Key Laboratory of Nuclear Physics and Technology, School of Physics,
Peking University, Beijing 100871, China
}
\affiliation{
Southern Center for Nuclear-Science Theory (SCNT), Institute of Modern Physics, Chinese Academy of Sciences, Huizhou 516000,  China
}

\begin{abstract}
To synthesize new superheavy elements, the accurate prediction of nuclear masses of superheavy nuclei is essential
for
calculations of  reaction $Q$ values, neutron separation energies and $\alpha$-decay energies, which are important
for estimating beam energies, survival probabilities and also for identifications. 
In this work, we include existing $\alpha$-decay energies of superheavy nuclei in the fitting procedure of
extended Skyrme density functionals as corresponding nuclear masses  are not available.
Systematic $\alpha$-decay energies are well reproduced with deviations smaller than 0.2 MeV.
The high quality $\alpha$-decay energies make it feasible for direct identification of new elements and new isotopes.
The resulting binding energies in the heaviest region are surprisingly close to the inferences by AME2020.
Our work should be useful for guiding experimental synthesis of new elements 119 and 120. 

\end{abstract}
\maketitle
\section{introduction}
To synthesize new elements in the 8th row of the periodic table is now one of the major scientific problems.
 To date, superheavy elements up to Z=118 have
been synthesized~\cite{118}. 
Experimental attempts to synthesize
Z = 119 and 120 elements were performed in laboratories using reactions such as $^{58}$Fe + $^{244}$Pu ~\cite{Fe58} at JINR, $^{51}$V + $^{248}$Cm
~\cite{V51} at RIKEN, and $^{64}$Ni + $^{238}$U, $^{50}$Ti + $^{249}$Bk, $^{50}$Ti + $^{249}$Cf,
$^{54}$Cr + $^{248}$Cm ~\cite{gsi119,gsi120} at GSI, but no evidence of new elements was observed yet.
This is now a highly competitive field and reliable theoretical   guidance is very desirable for such extremely difficult experiments.

For experiments, nuclear masses are important for guiding the synthesis of superheavy nuclei in several aspects.
The reaction $Q$ values and excitation energy of compound nuclei are important in calculations of production cross sections~\cite{Zagrebaev,gsi119,zhulong,dengxq,fszhang}.
The neutron separation energies $S_n$ are important for calculations of survival probabilities of compound nuclei after multiple neutron evaporations
and thus optimal bombarding energies~\cite{Zagrebaev,zubov,qiao}.  
The $\alpha$-decay  energies are important for identifying new elements and $\alpha$-decay chains~\cite{118,renzz,nazarewicz}. 
However, experimental binding energies are only available up to Z=110 nuclei~\cite{wangm}.
Fortunately,  $\alpha$-decay  energies are available up to Z=118, providing
precious constrains on theories. 

In theoretical estimations of cross sections, nuclear masses from FRDM model are
widely adopted~\cite{frdm}, which has high precision compared with available experimental data. 
Besides, machine learning of nuclear masses and other mass models
are very precise for existing data~\cite{low,niu,ws4,ymzhao,nazarewicz2,utama}. 
However, the extrapolation to superheavies is difficult since superheavy nuclei with strong repulsive Coulomb potential 
can redistribute density distributions and shell structures, so that exotic bubble or semi-bubble structures are possible~\cite{nazarewicz3,jcpei,Pomorski,bulgac}. 
In this respect, self-consistent microscopic theories have powerful predictive capabilities
for calculating nuclear masses of superheavy nuclei.
It has been a longstanding issue that the precision of self-consistent microscopic theory 
for nuclear masses
is not very satisfactory. 
In this work, we aim to optimize the Skyrme energy density functional theory (DFT)
especially for superheavy nuclei, in which experimental $\alpha$-decay  energies are included in the fitting. 
It is known that nuclear DFT is not precise for light nuclei as quantum corrections play a significant role~\cite{bender}. 
Actually superheavy nuclei are similar to mesoscopic systems and DFT should be a very suitable tool. 
As a result, we obtain high quality microscopic nuclear masses of superheavy nuclei.

\begin{figure*}[t]
\centering
\includegraphics[width=0.9\textwidth]{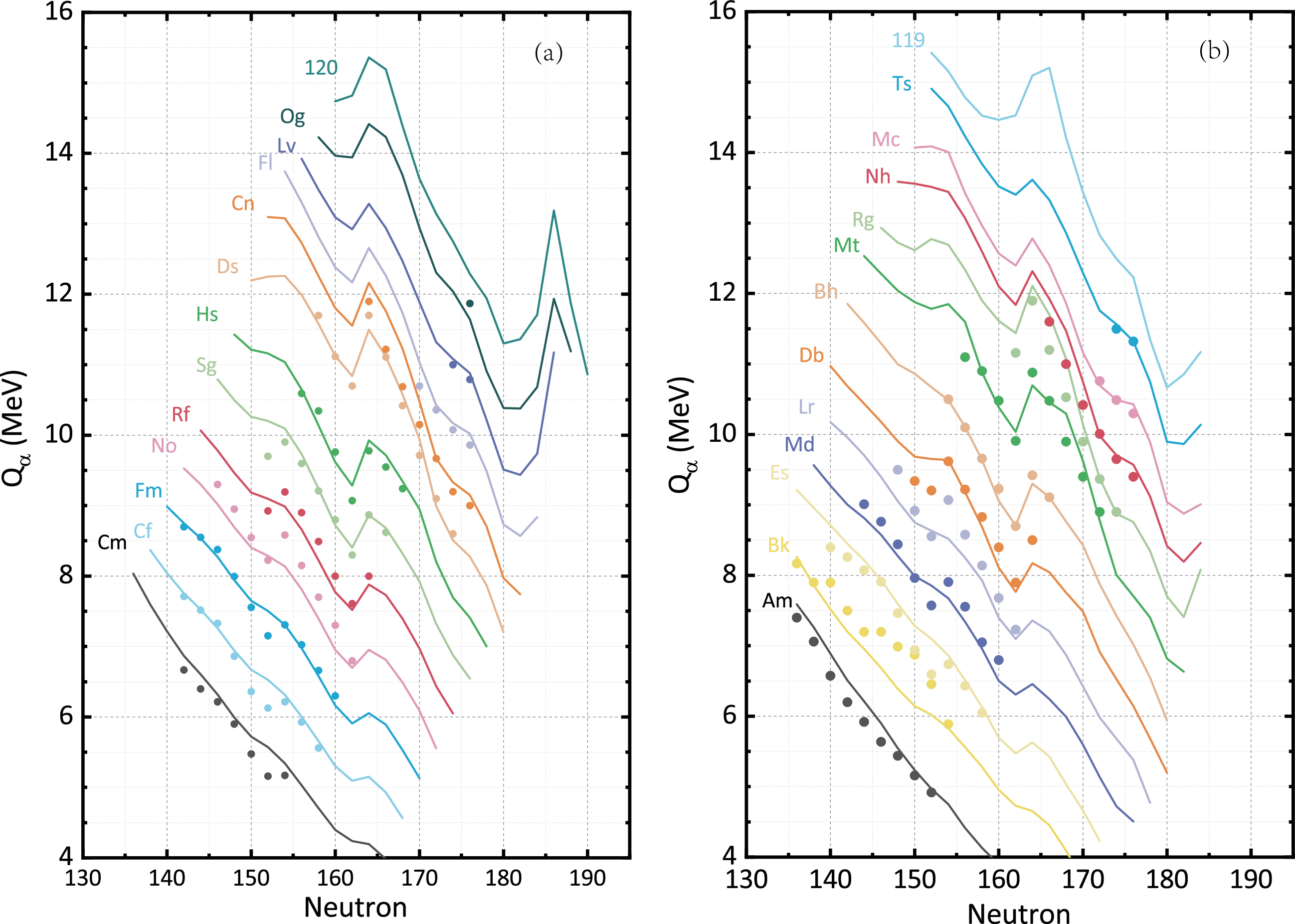}
\caption{
Systematically calculated $Q_{\alpha}$ of heavy and superheavy nuclei from Am to Z=120, using the fitted extended Skyrme parameterizations in this work. 
The experimental data~\cite{wangm} are given in dots. 
                           \label{Fig1}
}
\end{figure*}

\begin{figure*}[ht]
\centering
\includegraphics[width=0.96\textwidth]{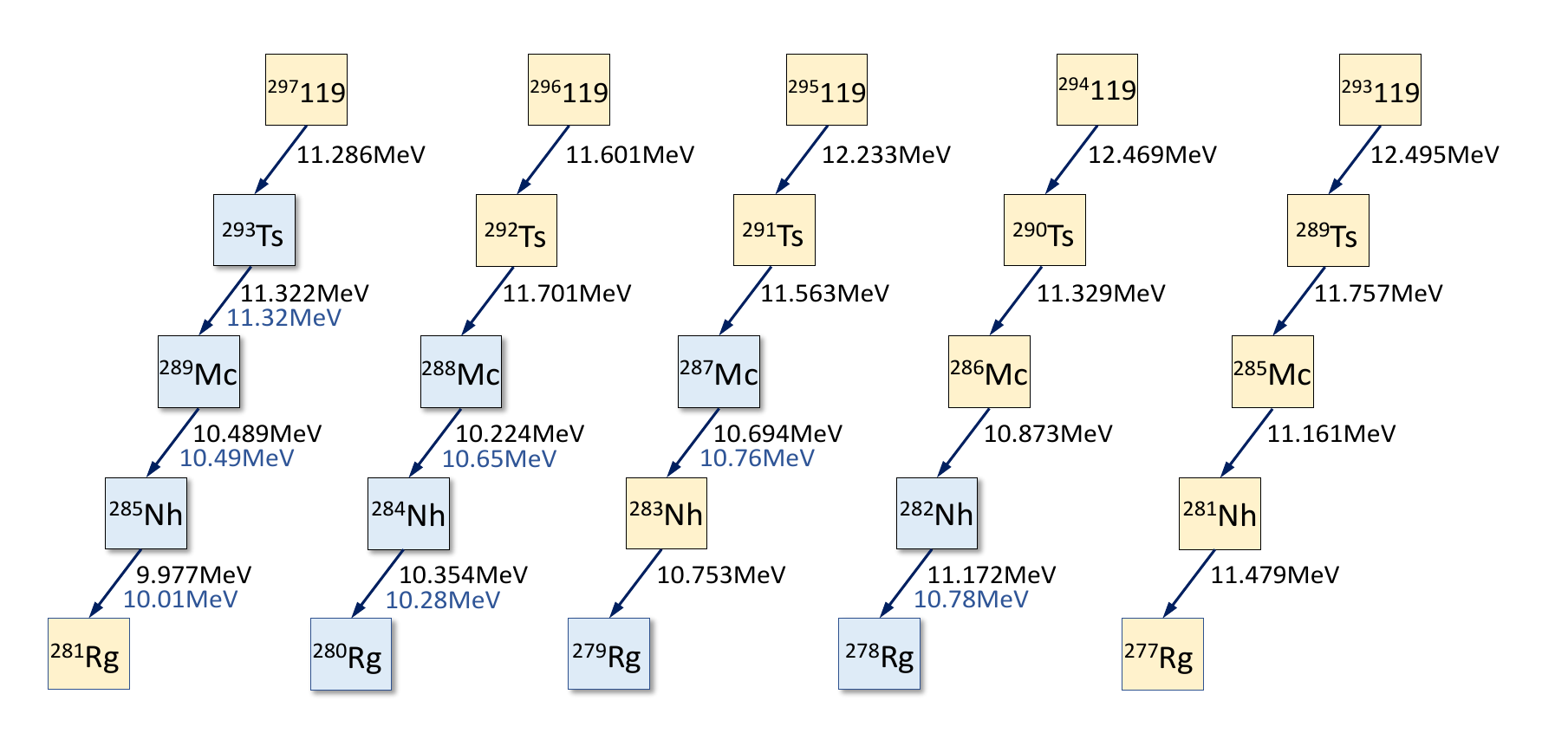}
\caption{
Calculated  $Q_{\alpha}$ of possible $\alpha$-decay chains from $Z$=119 nuclei in future experiments.
The experimental known data~\cite{wangm} are also given in blue colour. 
                           \label{Fig3}
}
\end{figure*}

\section{methods}
In this work we performed new optimizations of extended Skyrme forces by including  $\alpha$-decay energies of sueprheavy nuclei up to $Z$=118.
The extension of Skyrme forces with an additional higher-order density dependent term is expected to provide
better descriptions of finite nuclei and nuclear matter within a large range of densities~\cite{xiong}. 
The Skyrme interaction  includes a two-body interaction ${v_{ij}^{(2)}}$ and a three-body interaction $v_{ijk}^{(3)}$.
The low-momentum effective two-body interaction can be written as~\cite{sly4},
\begin{equation}
\begin{array}{ll}
{v_{ij}^{(2)}}=&\displaystyle t_{0}(1+x_{0}P_{\sigma})\delta({\rr}_i-{\rr}_j)\vspace{5pt}\\
&+\displaystyle \frac{1}{2}t_{1}(1+x_1P_{\sigma})[\delta({\rr}_i-{\rr}_j){\bf{k}}^2+{\bf{k'}}^2\delta({\rr}_i-{\rr}_j)] \vspace{5pt}\\
&+\displaystyle t_2(1+x_2P_{\sigma}){\bf{k'}}\cdot\delta({\rr}_i-{\rr}_j){\bf{k}} \vspace{5pt}\\
&+\displaystyle iW_0(\sigma_i+\sigma_j)\cdot{\bf{k'}}\times\delta({\rr}_i-{\rr}_j){\bf{k}}
\end{array}
\label{eqn.02}
\end{equation}
The three-body interaction can be transformed into a density dependent two-body interaction, and
an additional term is adopted in this work~\cite{xiong}. 
\begin{equation}
\begin{array}{llll}
    v_{ijk}^{(3)}
     =&\displaystyle \frac{1}{6}t_3(1+x_3P_\sigma)\rho(\boldsymbol{R})^\gamma\delta(\rr_i-\rr_j) \\
      &\displaystyle +\frac{1}{6}t_{3E}(1+x_{3E}P_{\sigma})\rho(\boldsymbol{R})^{\gamma_E}\delta(\rr_i-\rr_j).
                                                                \label{eqn.03}
\end{array}
\end{equation}
In Eqs.(\ref{eqn.02}, \ref{eqn.03}), $t_i$, $x_i$ and $W_0$ are parameters of the Skyrme interation. The spin-orbit term can be
extended to include isospin dependence and $W_0$ is replaced by $b_4$ and $b_4'$~\cite{so}.
Besides, $t_{3E}$ and $x_{3E}$ are additional high-order parameters. 
The power factor $\gamma$ takes $1/6$, which is the same as SLy4~\cite{sly4},  and the high-order power $\gamma_E$ takes $1/2$~\cite{xiong}. 
The effective mass is taken as $m^{*}/m$=0.8, while it is  0.68 for SLy4~\cite{sly4}. 
In our work, systematic calculations are based on the self-consistent deformed Skyrme-Hartree-Fock+BCS (SHF-BCS) method~\cite{skyax}.
In addition, the two-body center-of-mass correction is included after variations~\cite{com}, which has been studied systematically before~\cite{zuo}.
The mixed pairing interaction~\cite{mix} is adopted and the pairing strengths are $V_p$=480 MeV and $V_n$=440 MeV.
The Hartree-Fock-BCS equations are solved by the SKYAX code in axial-symmetric coordinate-spaces~\cite{skyax}.

\begin{table}
\caption{\label{table1}
The refitted Skyrme parameters (Sk-SHE1) in this work by including $Q_{\alpha}$ of superheavy nuclei. The incompressibility $K_\infty$ (MeV) and symmetry energy  $a_s$ (MeV)
at saturation density are also given.  }
\begin{ruledtabular}
\begin{tabular*}{\textwidth}{@{\extracolsep{\fill}}cccc}
Parameters  & Values   & Parameters  & Values    \\
\hline
$t_0$       & -2352.7014 & $x_0$      & 0.3103       \\
$t_1$       & 387.8619   & $x_1$      & -0.6574      \\
$t_2$       & -133.2639  & $x_2$      & -0.04187     \\
$t_3$       & 11819.3660 & $x_3$      & 0.5810       \\
$t_{3E}$    & 2608.4635  & $x_{3E}$   &-1.0028       \\
$b_4$       & 61.8862    & $b_4^{'}$  & 43.0657      \\
$\gamma$    & 1/6      & $\gamma_E$ &0.5         \\
$\rho_0$    & 0.1596    & $e_\infty$ &-15.9953      \\
$K_\infty$  &233.2128    & $a_s$      & 32.3017      \\
\end{tabular*}
\end{ruledtabular}
\label{table1}
\end{table}

\begin{table}[]
\caption{\label{table2}
Calculated $Q_{\alpha}$ (MeV) in the heaviest region in this work with comparisons with experimental data~\cite{wangm}, and predictions
by FRDM2012~\cite{frdm} and WS4~\cite{ws4} models. }
\begin{ruledtabular}
\begin{tabular*}{\textwidth}{@{\extracolsep{\fill}}ccccc}
 {}                 & Present & FRDM2012 & WS4 & Expt. \\
\hline
$^{294}_{118}Og$    & 11.642         &  12.366         & 12.199	        &  11.87 \\
$^{293}_{117}Ts$    & 11.322         &  11.396         & 11.622	        &  11.32 \\
$^{291}_{117}Ts$    & 11.563         &  11.746         & 11.719	        &  11.5 \\
$^{292}_{116}Lv$    & 10.881         &  10.816         & 11.127	        &  10.791 \\
$^{290}_{116}Lv$    & 11.080         &  11.056         & 11.085	        &  11 \\
$^{291}_{115}Mc$    & 10.425         &  9.726          & 10.192	        &  10.3 \\
$^{289}_{115}Mc$    & 10.489         &  10.096         & 10.296	        &  10.49 \\
$^{287}_{115}Mc$    & 10.694         &  10.216         & 10.502	        &  10.76 \\
\end{tabular*}
\end{ruledtabular}
\label{table2}
\end{table}

In the fitting procedure, we minimize the quantity
\begin{equation}
\begin{array}{ll}
  \chi^2 =&\displaystyle (\frac{e_\infty+16.0}{0.2})^2+(\frac{\rho_s-0.16}{0.005})^2 \vspace{5pt}\\
          &\displaystyle+\sum_i(\frac{\sqrt{<r^2>_{\rm ch}(i)}-\sqrt{<r^2>_{\rm ch}^{\rm exp}(i)}}{0.02})^2 \vspace{4pt}\\
          &\displaystyle+\sum_i\frac{(Q_{\alpha}(i)-Q_{\alpha}^{\rm exp}(i))^2}{0.004}+\sum_i(\frac{E(i)-E^{\rm exp}(i)}{2.0})^2.
\end{array}
\label{eq04}
\end{equation}
where $e_\infty$ is the average energy per nucleon about -16.0 MeV at the saturation density and the saturation density $\rho_s$ is constrained to be around 0.16 fm$^{-3}$; 
$E^{\rm exp}(i)$ and $\sqrt{<r^2>_{\rm ch}^{\rm exp}(i)}$ denote binding energies and charge radii of selected nuclei, as listed in the Appendix.
The nuclear matter properties such as incompressibility and symmetry energy at $\rho_s$ are not strictly constrained in the fitting. 
The optimization is realized by the simulated annealing method via gradually decreasing temperatures, which as been applied in Refs.~\cite{xiong,agrawal}.
We adjust $t_2$, $t_3$, $x_2$, $x_3$, $x_{3E}$, $b_4$, $b_4^{'}$ and determine the rest parameters are determined by relations in
the equation of state. The obtained Skyrme parameters are given in the Table \ref{table1}.

\section{Results}

Firstly the calculated $\alpha$-decay energies in the heavy and superheavy region are shown from Am to $Z$=120.
The results of even-$Z$ nuclei are shown in Fig.\ref{Fig1}(a) and odd-$Z$ are shown in Fig.\ref{Fig1}(b).
It can be seen that the systematics of $Q_{\alpha}$ are nicely reproduced. 
The deformed shell effects around $N$=162 are also exhibited in $Q_{\alpha}$, but the deformed shell effects around $N$=152
are much underestimated. 
Note that only $Q_{\alpha}$ of even-even nuclei are included in the fitting.
However, the systematics of $Q_{\alpha}$ are also well reproduced for odd-$Z$ nuclei. 
In particular, the systematics above $Z$=110 are very well described.
Table \ref{table2} displays $Q_{\alpha}$ in the heaviest region calculated by different models compared with experiments.
Our results agree with experimental data~\cite{wangm} within 0.2 MeV, while the largest deviations of FRDM2012~\cite{frdm} and WS4~\cite{ws4} are around 0.5 MeV and 0.3 MeV, respectively. 
This means  $Q_{\alpha}$ of new superheavy elements or new isotopes can be reliably predicted.
Several microscopic calculations of  $Q_{\alpha}$ of even-even nuclei are presented in Ref.~\cite{nazarewicz},
in which the best average standard deviation is around 0.3 MeV. 
Some models such as SLy4 are not good at descriptions of binding energies but rather good at  $Q_{\alpha}$.
The models that are good at binding energies don't promise good descriptions of $Q_{\alpha}$ values. 
For example, $Q_{\alpha}$ of $^{294}$Og is much underestimated by UNEDF0 for more than 1.5 MeV~\cite{unedf0}.

Currently there are strong interests in experimental synthesis of next superheavy element 119 via $^{50}$Ti+$^{249}$Bk~\cite{gsi119} and $^{54}$Cr+$^{243}$Am~\cite{Cr54} reactions. 
Fig.\ref{Fig3} displays the possible $\alpha$-decay chains after evaporations of 2 to 4 neutrons from primary $Z$=119 compound nuclei.
It can be seen that the existing experimental data can be well reproduced.
For even-$N$ nuclei, the experimental $Q_{\alpha}$ is reproduced perfectly within 0.1 MeV. 
However, for odd-odd nuclei, the deviations in $Q_{\alpha}$ are larger and the largest deviation is about 0.4 MeV.
This is because $Q_{\alpha}$ values in odd-odd nuclei are sensitive to orbital structures. 
$Q_{\alpha}$ values of 119 isotopes by UNEDF0 are significantly smaller than our results, as $Q_{\alpha}$ of $^{294}$Og is also much underestimated by UNEDF0~\cite{unedf0}. 
Our high quality microscopic results of  $Q_{\alpha}$ are due to the inclusion of experimental  $Q_{\alpha}$ in the fitting
, which provides
a stepping stone for reliable inferences of  $Q_{\alpha}$ of 119 and 120 elements.  
We hope our accurate  $Q_{\alpha}$ values are useful for direct identifications of new elements or new isotopes. 

\begin{figure}[ht]
\centering
\includegraphics[width=0.49\textwidth]{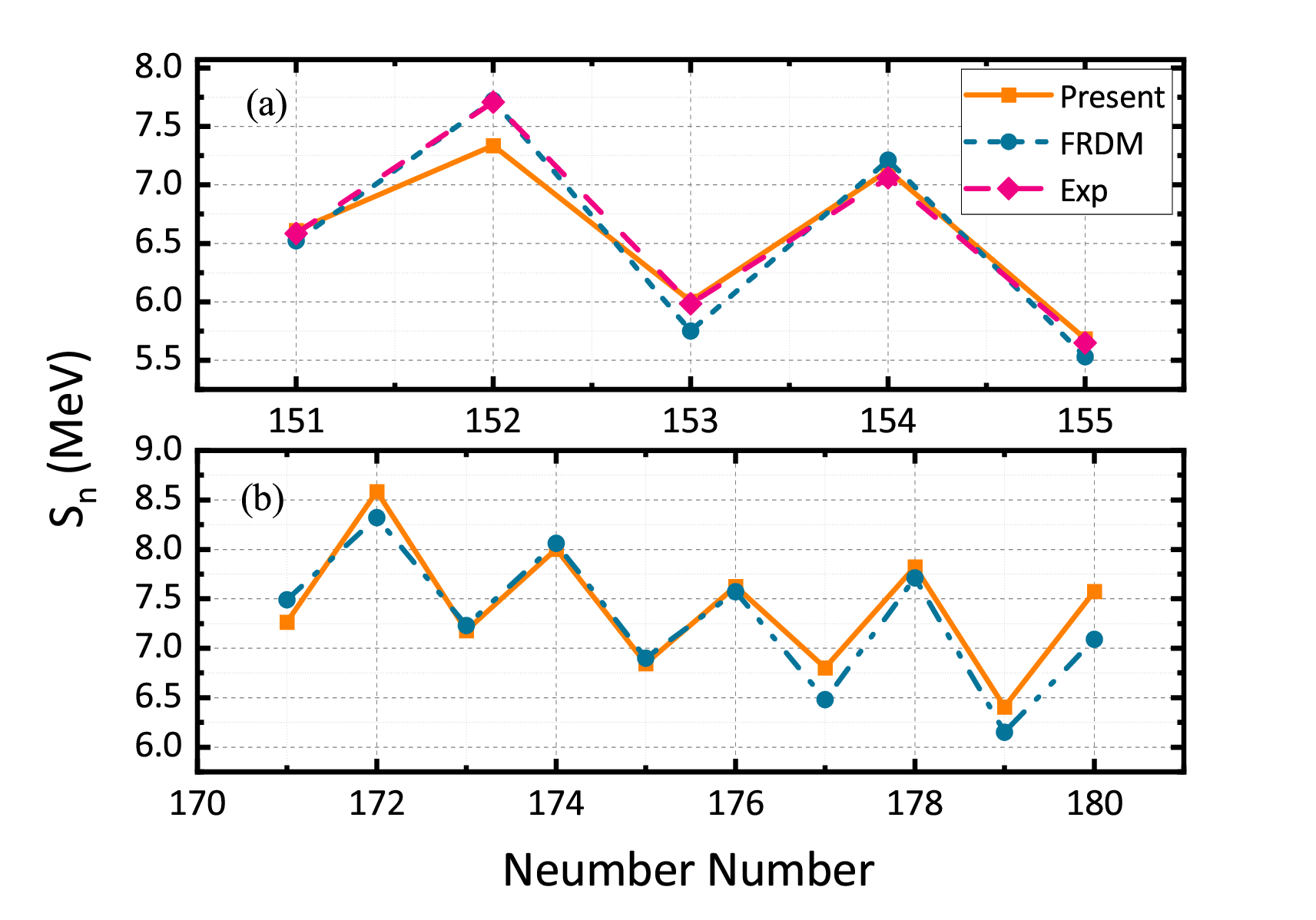}
\caption{
Calculated neutron separation energies $S_n$ of No  and $Z$=119 isotopes
with comparison with experimental data and FRDM2012 results.
                           \label{Fig4}
}
\end{figure}


In calculations of survival probabilities of highly excited compound superheavy nuclei after evaporations of
multiple neutrons, the accurate calculations of neutron separation energies are important. 
Fig.\ref{Fig4} shows results of neutron separation energies ($S_n$) of No isotopes with experimental data and $Z$=119 isotopes.
For comparison, $S_n$ from FRDM mass model are also shown. 
For No isotopes, $S_n$ can be well reproduced except for $N$=152, where the deformed shell at $N$=152 is underestimated. 
This means the addopted pairing strength is reasonable for the superheavy region. 
For $Z$=119 isotopes, the two models are very close except for $N$=180. 
The $S_n$ in our calculations at $N$=180 is 7.57 MeV, which is larger than 7.12 MeV of FRDM2012~\cite{frdm} by 0.45 MeV, due to the shell effects around $N$=184.
It is known that spherical shell effects in Skyrme density functionals are usually overestimated as the effective mass is smaller than
1~\cite{xiong}. 
Besides, $S_n$ from WS4 model is close to that of FRDM2012 at $N$=180. $S_n$ at $N$=180 by UNEDF0 is about 7.27 MeV. Thus it is likely that $S_n$ is slightly overestimated by our calculations at $N$=180. 
The production cross sections of superheavy nuclei are very much dependent on fission of compound nuclei, which 
will be studied in a forthcoming work. 

The systematically calculated binding energies of superheavy nuclei are shown in Table~\ref{table5} in the Appendix.
For comparison, the binding energies from FRDM2012~\cite{frdm}, WS4~\cite{ws4} and AME2020~\cite{wangm} are also listed. 
In the superheavy region, it can be seen that our results are systematically smaller than experiments about 0.6 MeV.
However, the trends  of systematics are rather stable. 
Generally, our results are surprisingly close to AME2020 evaluations. 
The FRDM2012 binding energies are overestimated by $\sim$4 MeV compared to AME2020  in the heaviest region. 
The WS4 binding energies are generally larger than AME2020 by 1$\sim$2 MeV in the heaviest region. 
There are already very good descriptions of global nuclear masses by UNEDF0~\cite{unedf0}, which are also close to AME2020. 
Based on the binding energies, the reaction $Q$ values can be calculated.
For $^{50}$Ti+$^{249}$Bk and $^{54}$Cr+$^{243}$Am reactions, the reaction $Q$ values are -192.381 and  -208.395 MeV, respectively.
Considering the systematical underestimation of binding energies about 0.6 MeV, $Q$ values should be around -191.78 and -207.79 MeV, respectively. 
The corrected $Q$ values are close to that of WS4 model, while it is overestimated by FRDM2012 about 4 MeV.
This is important for calculating  excitation energies of compound nuclei and beam energies.
In the future, the direct mass measurements of superheavy nuclei are expected to be very useful~\cite{gates}.

\section{summary}

We have optimized the extended Skyrme density functional in particular for the superheavy mass region. 
Compared to existing parameterizations, experimental $Q_{\alpha}$ of even-even superheavy nuclei up to $Z$=118 are 
included in the fitting. As a result, $Q_{\alpha}$ of all existing superheavy nuclei can be very well reproduced, while
the discrepancies are slightly larger for odd-odd superheavy nuclei. 
The precise prediction of  $Q_{\alpha}$ should be helpful for direct identification of new superheavy elements or new isotopes. 
The obtained binding energies are also very satisfactory, which are surprisingly close to AME2020 inferences. 
Neutron separation energies and reaction $Q$ values are also studied in details aiming the synthesis of new element 119. 
Thanks to the powerful supercomputing capabilities, 
the inferences of masses of superheavy nuclei have made great progress in recent years. 
We expect the present high quality microscopic nuclear masses in the heaviest region are useful for guiding experimental synthesis of superheavvy nuclei.

\acknowledgments
We are grateful for  discussions in the workshop on ``machine learning in synthesis of superheavy nuclei" in Huizhou in August 23th, 2023.
This work was supported by the National Key R$\&$D Program of China (Grant No. 2023YFE0101500,2023YFA1606403)
 and National Natural Science Foundation of China under Grants
No.  12335007 and 11961141003..
We also acknowledge the funding support from the State Key Laboratory of Nuclear Physics and Technology, Peking University (No. NPT2023ZX01).

\appendix

\begin{table*}
\caption{Calculated binding energies (MeV) of superheavy nuclei, with comparison with AME2020 evaluations, FRDM2012, and WS4 models. } 
\begin{ruledtabular}
\begin{tabular*}{\textwidth}{@{\extracolsep{\fill}}cccccc}
 Z  &N & Present & FRDM2012 & WS4 & AME2020 \\
\hline
110	&	157	&	-1933.996	&	-1936.12	&	-1934.4471	&	1935.216	(\#)	\\
110	&	158	&	-1942.618	&	-1944.79	&	-1943.1411	&	1943.536	(\#)	\\
110	&	159	&	-1949.511	&	-1951.94	&	-1950.1299	&	1950.29304	\\	
110	&	160	&	-1958.001	&	-1960.4	    &	-1958.6971	&	1958.5152	\\	
110	&	161	&	-1964.840	&	-1967.33	&	-1965.5894	&	1965.292	(\#)	\\
110	&	162	&	-1972.780	&	-1975.56	&	-1973.8763	&	1973.36	(\#)	\\
110	&	163	&	-1978.782	&	-1981.96	&	-1980.1013	&	1979.25	(\#)	\\
110	&	164	&	-1985.995	&	-1989.4  	&	-1987.6934	&	1986.226	(\#)	\\
110	&	165	&   -1991.582	&	-1995.01	&	-1993.3555	&	1991.825	(\#)	\\
110	&	166	&	-1998.668	&	-2001.98	&	-2000.4899	&	1999.068	(\#)	\\
110	&	167	&	-2004.127	&	-2007.44	&	-2006.0375	&	2004.649	(\#)	\\
110	&	168	&	-2010.957	&	-2014.24	&	-2013.0286	&	2011.608	(\#)	\\
110	&	169	&	-2016.205	&	-2020.04	&	-2018.5295	&	2016.891	(\#)	\\
110	&	170	&	-2022.996	&	-2027.16	&	-2025.5051	&	2023.56	(\#)	\\
110	&	171	&	-2028.193	&	-2032.83	&	-2030.8438	&	2028.82	(\#)	\\
110	&	172	&	-2035.035	&	-2039.68	&	-2037.6613	&	2035.194	(\#)	\\
110	&	173	&	-2040.384	&	-2045.17	&	-2042.9587	&	2040.43	(\#)	\\
110	&	174	&	-2046.708	&	-2051.83	&	-2049.6333	&	2046.788	(\#)	\\
111	&	161	&	-1965.071	&	-1967.58	&	-1965.5933	&	1965.744	(\#)	\\
111	&	162	&	-1973.226	&	-1975.85	&	-1973.8625	&	1973.79	(\#)	\\
111	&	163	&	-1979.556	&	-1982.66	&	-1980.5345	&	1980.198	(\#)	\\
111	&	164	&	-1986.930	&	-1990.07	&	-1988.1814	&	1987.425	(\#)	\\
111	&	165	&	-1992.940	&	-1996.16	&	-1994.3326	&	1993.548	(\#)	\\
111	&	166	&	-2000.068	&	-2003.26	&	-2001.595	&	2000.494	(\#)	\\
111	&	167	&	-2005.782	&	-2009.13	&	-2007.5721	&	2006.604	(\#)	\\
111	&	168	&	-2012.870	&	-2016.14	&	-2014.703	&	2013.264	(\#)	\\
111	&	169	&	-2018.320	&	-2022.35	&	-2020.6642	&	2019.36	(\#)	\\
111	&	170	&	-2025.342	&	-2029.49	&	-2027.6208	&	2025.729	(\#)	\\
111	&	171	&	-2030.762	&	-2035.51	&	-2033.333	&	2031.528	(\#)	\\
111	&	172	&	-2037.770	&	-2042.45	&	-2040.1891	&	2037.883	(\#)	\\
111	&	173	&	-2043.283	&	-2048.29	&	-2045.8768	&	2043.38	(\#)	\\
111	&	174	&	-2049.756	&	-2055	    &	-2052.5217	&	2049.72	(\#)	\\
111	&	175	&	-2055.032	&	-2060.56	&	-2057.8518	&	2054.91	(\#)	\\
112	&	164	&	-1988.916 	&	-1991.96	&	-1990.2504	&	1989.684	(\#)	\\
112	&	165	&	-1995.115	&	-1998.13	&	-1996.4717	&	1995.785	(\#)	\\
112	&	166	&	-2002.530	&	-2005.54	&	-2004.2101	&	2003.268	(\#)	\\
112	&	167	&	-2008.476	&	-2011.65	&	-2010.2321	&	2009.358	(\#)	\\
112	&	168	&	-2015.739	&	-2019.2 	&	-2017.923	&	2016.56	(\#)	\\
112	&	169	&	-2021.486	&	-2025.5 	&	-2023.8483	&	2022.357	(\#)	\\
112	&	170	&	-2028.774  	&	-2033.11	&	-2031.184	&	2029.554	(\#)	\\
112	&	171	&	-2034.411	&	-2039.21	&	-2036.9823	&	2035.336	(\#)	\\
112	&	172	&	-2041.626	&	-2046.49	&	-2044.257	&	2042.244	(\#)	\\
112	&	173	&	-2047.353	&	-2052.34	&	-2049.9077	&	2047.725	(\#)	\\
112	&	174	&	-2054.002	&	-2059.36	&	-2056.9194	&	2054.338	(\#)	\\
112	&	175	&	-2059.485	&	-2064.93	&	-2062.1919	&	2059.512	(\#)	\\
112	&	176	&	-2065.856	&	-2071.54	&	-2068.822	&	2066.112	(\#)	\\
113	&	165	&	-1995.663	&	-1998.44	&	-1996.5884	&	1996.318	(\#)	\\
113	&	166	&	-2003.302	&	-2006.5	    &	-2004.337	&	2004.057	(\#)	\\
113	&	167	&	-2009.459	&	-2013.01	&	-2010.8184	&	2010.4	(\#)	\\
113	&	168	&	-2016.894  	&	-2020.81	&	-2018.6071	&	2017.861	(\#)	\\
113	&	169	&	-2022.906	&	-2027.42	&	-2024.9623	&	2023.914	(\#)	\\
113	&	170	&	-2030.422   	&	-2034.99	&	-2032.5908	&	2031.091	(\#)	\\
113	&	171	&	-2036.262	&	-2041.39	&	-2038.845	&	2037.132	(\#)	\\
113	&	172	&	-2043.656	&	-2048.67	&	-2046.1063	&	2044.02	(\#)	\\
113	&	173	&	-2049.558	&	-2054.84	&	-2052.1379	&	2050.048	(\#)	\\
113	&	174	&	-2056.361	&	-2061.86	&	-2059.1401	&	2056.642	(\#)	\\
113	&	175	&	-2061.992	&	-2067.75	&	-2064.8269	&	2062.08	(\#)	\\
113	&	176	&	-2068.481	&	-2074.49	&	-2071.4983	&	2068.662	(\#)	\\
113	&	177	&	-2073.864	&	-2080.25	&	-2076.9885	&	2074.08	(\#)	\\
\end{tabular*}
\end{ruledtabular}
\label{table5}
\end{table*}

\begin{table*}
\begin{ruledtabular}
\begin{tabular*}{\textwidth}{@{\extracolsep{\fill}}cccccc}
 Z  &N &Present & FRDM2012 & WS4 & AME2020 \\
\hline
114	&	170	&	-2033.000	&	-2037.98	&	-2035.6461	&	2034.292	(\#)	\\
114	&	171	&	-2039.038	&	-2044.36	&	-2041.8638	&	2040.315	(\#)	\\
114	&	172	&	-2046.647	&	-2051.93	&	-2049.51	&	2047.474	(\#)	\\
114	&	173	&	-2052.771	&	-2058.11	&	-2055.5078	&	2053.485	(\#)	\\
114	&	174	&	-2059.756	&	-2065.62	&	-2062.9072	&	2060.352	(\#)	\\
114	&	175	&	-2065.595	&	-2071.58	&	-2068.5979	&	2066.061	(\#)	\\
114	&	176	&	-2072.280	&	-2078.82	&	-2075.6951	&	2072.63	(\#)	\\
114	&	177	&	-2077.876	&	-2084.47	&	-2081.2188	&	2078.031	(\#)	\\
115	&	172	&	-2048.024	&	-2053.07	&	-2050.385	&	2048.893	(\#)	\\
115	&	173	&	-2054.334	&	-2059.55	&	-2056.7437	&	2054.88	(\#)	\\
115	&	174	&	-2061.463	&	-2066.87	&	-2064.1064	&	2062.015	(\#)	\\
115	&	175	&	-2067.457	&	-2073.1	    &	-2070.1501	&	2067.99	(\#)	\\
115	&	176	&	-2074.232	&	-2080.43	&	-2077.2439	&	2074.539	(\#)	\\
115	&	177	&	-2079.886	&	-2086.46	&	-2083.1909	&	2080.208	(\#)	\\
116	&	173	&	-2056.513	&	-2061.56	&	-2058.9841	&	2057.391	(\#)	\\
116	&	174	&	-2063.864	&	-2069.17	&	-2066.7212	&	2064.8	(\#)	\\
116	&	175	&	-2070.084	&	-2075.39	&	-2072.6829	&	2070.756	(\#)	\\
116	&	176	&	-2077.171	&	-2083.1 	&	-2080.0764	&	2077.872	(\#)	\\
116	&	177	&	-2083.154   &	-2089.1 	&	-2086.1008	&	2083.523	(\#)	\\
117	&	174	&	-2064.758 	&	-2069.62	&	-2066.9617	&	2065.518	(\#)	\\
117	&	175	&	-2070.929	&	-2076.14	&	-2073.2869	&	2071.74	(\#)	\\
117	&	176	&	-2078.436 	&	-2083.77	&	-2080.7803	&	2078.835	(\#)	\\
117	&	177	&	-2084.568	&	-2090.11	&	-2087.0693	&	2085.048	(\#)	\\
118	&	175	&	-2072.815	&	-2077.52	&	-2075.0381	&	2073.854	(\#)	\\
118	&	176	&	-2080.518	&	-2085.1 	&	-2082.8186	&	2081.226	(\#)	\\
118	&	177	&	-2086.899	&	-2091.77	&	-2089.0759	&	2087.42	(\#)	\\
119	&	174	&	-2066.370	&	-2070.51	&	-2068.063	&			\\
119	&	175	&	-2073.123	&	-2077.41	&	-2074.7834	&			\\
119	&	176	&	-2080.821	&	-2084.98	&	-2082.499	&			\\
119	&	177	&	-2087.623	&	-2091.46	&	-2089.1077	&			\\
119	&	178	&	-2095.446 	&	-2099.17	&	-2096.6519	&			\\
119	&	179	&	-2101.854	&	-2105.32	&	-2102.6516	&			\\
119	&	180	&	-2109.429	&	-2112.41	&	-2109.7751	&			\\
119	&	181	&	-2115.578  	&	-2118.38	&	-2115.7083	&			\\
119	&	182	&	-2122.734	&	-2125.1	    &	-2122.6296	&			\\
119	&	183	&	-2128.614	&	-2130.8	    &	-2128.2368	&			\\
119	&	184	&	-2135.202	&	-2137.2  	&	-2134.7913	&			\\
120	&	176	&	-2082.245	&	-2085.7 	&	-2083.6599	&			\\
120	&	177	&	-2088.936	&	-2092.17	&	-2090.1904	&			\\
120	&	178	&	-2096.862	&	-2100.16	&	-2098.1074	&			\\
120	&	179	&	-2103.110	&	-2106.33	&	-2104.1125	&			\\
120	&	180	&	-2110.237	&	-2113.73	&	-2111.5974	&			\\
120	&	181	&	-2117.857	&	-2119.7	    &	-2117.5237	&			\\
120	&	182	&	-2125.210	&	-2126.77	&	-2124.8457	&			\\
120	&	183	&	-2131.218	&	-2132.47	&	-2130.4778	&			\\
120	&	184	&	-2137.956	&	-2139.18	&	-2137.4419	&			\\
\end{tabular*}
\end{ruledtabular}
\label{table5}
\end{table*}

\begin{table}
\caption{\label{table6}
The selected nuclei used for fitting,  in which binding energies (E$_{B}$), charge radii ($R_c$) and Q$_\alpha$ of some nuclei are employed.}
\begin{ruledtabular}
\begin{tabular*}{0.49\textwidth}{@{\extracolsep{\fill}}ccccc}
Z   &   N   & E$_{B}$  & Q$_\alpha$ & $R_c$ \\
\hline
8	&	8	&	$\surd$	&	&$\surd$		\\
10	&	20	&	$\surd$	&		\\
12	&	12	&	$\surd$	&		\\
12	&	24	&	$\surd$	&		\\
14	&	28	&	$\surd$	&		\\
20	&	20	&	$\surd$	&	&$\surd$		\\
20	&	28	&	$\surd$	&	& $\surd$		\\
22	&	20	&	$\surd$	&		\\
24	&	26	&	$\surd$	&		\\
26	&	40	&	$\surd$	&		\\
26	&	44	&	$\surd$	&		\\
28	&	28	&	$\surd$	&	& $\surd$		\\
28	&	44	&	$\surd$	&		\\
28	&	50	&	$\surd$	&		\\
34	&	34	&	$\surd$	&		\\
38	&	64	&	$\surd$	&		\\
40	&	42	&	$\surd$	&		\\
40	&	60	&	$\surd$	&		\\
48	&	50	&	$\surd$	&		\\
48	&	78	&	$\surd$	&		\\
48	&	82	&	$\surd$	&		\\
50	&	70	&	$\surd$	&	&$\surd$		\\
50	&	82	&	$\surd$	&		\\
56	&	92	&	$\surd$	&		\\
60	&	96	&	$\surd$	&		\\
62	&	74	&	$\surd$	&		\\
64	&	96	&	$\surd$	&	& $\surd$	\\
70	&	82	&	$\surd$	&		\\
72	&	104	&	$\surd$	&		\\
72	&	114	&	$\surd$	&		\\
74	&	108	&	$\surd$	&		\\
74	&	116	&	$\surd$	&		\\
76	&	100	&	$\surd$	&		\\
76	&	108	&	$\surd$	&		\\
78	&	104	&	$\surd$	&		\\
82	&	96	&	$\surd$	&		\\
82	&	116	&	$\surd$	&		\\
82	&	126	&	$\surd$	&	&$\surd$		\\
82	&	128	&	$\surd$	&		\\
88	&	142	&	$\surd$	&		\\
90	&	126	&	$\surd$	&		\\
90	&	144	&	$\surd$	&		\\
92	&	126	&	$\surd$	&		\\
92	&	144	&	$\surd$	&		\\
94	&	146	&	$\surd$	&		\\
94	&	152	&	$\surd$	&		\\
96	&	154	&	$\surd$	&		\\
98	&	150	&	$\surd$	&		\\
98	&	152	&	$\surd$	&		\\
98	&	154	&	$\surd$	&	$\surd$	\\
98	&	156	&	$\surd$	&	$\surd$	\\
100	&	148	&	$\surd$	&		\\
100	&	150	&	$\surd$	&		\\
100	&	152	&	$\surd$	&		\\
100	&	154	&	$\surd$	&		\\
100	&	156	&	$\surd$	&		\\
102	&	150	&	$\surd$	&	$\surd$	\\
102	&	152	&	$\surd$	&	$\surd$	\\
102	&	154	&	$\surd$	&	$\surd$	\\
104	&	152	&	$\surd$	&	$\surd$	\\
104	&	154	&	$\surd$	&	$\surd$	\\
106	&	154	&	$\surd$	&	$\surd$	\\
\end{tabular*}
\end{ruledtabular}
\label{table6}
\end{table}

\begin{table}
\begin{ruledtabular}
\begin{tabular*}{0.49\textwidth}{@{\extracolsep{\fill}}ccccc}
Z   &   N   & E$_{B}$  & Q$_\alpha$ & $R_c$ \\
\hline
106	&	156	&	$\surd$	&	$\surd$	\\
106	&	160	&		&	$\surd$	\\
108	&	156	&	$\surd$	&		\\
108	&	158	&	$\surd$	&	$\surd$	\\
108	&	162	&		&	$\surd$	\\
110	&	160	&	$\surd$	&	$\surd$	\\
112	&	170	&		&	$\surd$	\\
114	&	174	&		&	$\surd$	\\
114	&	176	&		&	$\surd$	\\
116	&	174	&		&	$\surd$	\\
116	&	176	&		&	$\surd$	\\
118	&	176	&		&	$\surd$	\\
\end{tabular*}
\end{ruledtabular}
\label{table6}
\end{table}

\end{document}